\documentclass[pra,aps,10pt,superscriptaddress,twocolumn,floatfix,footinbib]{revtex4-1}
\usepackage{graphicx,color}
\usepackage[nice]{nicefrac}
\usepackage{amsmath,amssymb,bm,bbm}
\usepackage{amsmath}
\usepackage{braket}

\usepackage[plainpages=false,pdfpagelabels,colorlinks=true,linkcolor=red,urlcolor=blue,citecolor=blue,pdftitle={},pdfauthor={},pdfdisplaydoctitle=true,pdfduplex=DuplexFlipLongEdge]{hyperref}

\definecolor{darkred}{rgb}{0.90,0.2,0.2}
\definecolor{darkgreen}{rgb}{0,0.60,.2}
\definecolor{darkblue}{rgb}{0.1,0.3,1}
\definecolor{grey}{cmyk}{0,0,0,0.25}
\definecolor{orange}{cmyk}{0,0.6,0.8,0}

\begin{document}

\title{Entanglement in many-body eigenstates of quantum-chaotic quadratic Hamiltonians}

\author{Patrycja  \L yd\.{z}ba}
\affiliation{Department of Theoretical Physics, J. Stefan Institute, SI-1000 Ljubljana, Slovenia}
\affiliation{Department of Theoretical Physics, Wroclaw University of Science and Technology, 50-370 Wroc{\l}aw, Poland}
\author{Marcos Rigol}
\affiliation{Department of Physics, The Pennsylvania State University, University Park, Pennsylvania 16802, USA}
\author{Lev Vidmar}
\affiliation{Department of Theoretical Physics, J. Stefan Institute, SI-1000 Ljubljana, Slovenia}
\affiliation{Department of Physics, Faculty of Mathematics and Physics, University of Ljubljana, SI-1000 Ljubljana, Slovenia}

\begin{abstract}
In a recent Letter [\href{https://doi.org/10.1103/PhysRevLett.125.180604}{Phys.~Rev.~Lett.~{\bf 125},~180604~(2020)}], we introduced a closed-form analytic expression for the average bipartite von Neumann entanglement entropy of many-body eigenstates of random quadratic Hamiltonians. Namely, of Hamiltonians whose single-particle eigenstates have random coefficients in the position basis. A paradigmatic Hamiltonian for which the expression is valid is the quadratic Sachdev-Ye-Kitaev (SYK2) model in its Dirac fermion formulation. Here we show that the applicability of our result is much broader. Most prominently, it is also relevant for local Hamiltonians such as the three-dimensional (3D) Anderson model at weak disorder. Moreover, it describes the average entanglement entropy in Hamiltonians without particle-number conservation, such as the SYK2 model in the Majorana fermion formulation and the 3D Anderson model with additional terms that break particle-number conservation. We extend our analysis to the average bipartite second R{\'e}nyi entanglement entropy of eigenstates of the same quadratic Hamiltonians, which is derived analytically and tested numerically. We conjecture that our results for the entanglement entropies of many-body eigenstates apply to quadratic Hamiltonians whose single-particle eigenstates exhibit quantum chaos, to which we refer as quantum-chaotic quadratic Hamiltonians.
\end{abstract}
\maketitle


\section{Introduction} \label{sec:Intro}

Many-body quantum chaos is a phenomenon that underpins our understanding of quantum ergodicity in many-body systems, and is studied in different fields of physics ranging from condensed matter systems~\cite{dalessio_kafri_16}, quantum gases and analog quantum simulators~\cite{dalessio_kafri_16, eisert_friesdorf_15, mori_ikeda_18}, to high-energy physics~\cite{swingle_18}. It is also intimately related to the eigenstate thermalization hypothesis (ETH)~\cite{deutsch_91, srednicki_94, rigol_dunjko_08, dalessio_kafri_16, deutsch_18}, which provides a sufficient criterion to explain thermalization of local observables in isolated quantum systems. Establishing more rigorous relations between the ETH and many-body quantum chaos is an active research direction~\cite{dalessio_kafri_16, vidmar_rigol_17, garrison_grover_18, dymarsky_lashkari_18, bertini_kos_18, mierzejewski_vidmar_20, pandey_claeys_20}.

Early studies of quantum chaos focused on single-particle quantum systems that exhibit chaotic dynamics in the semiclassical limit~\cite{berry_77}. In this context, it was conjectured~\cite{bohigas_giannoni_84} that the statistics of the energy level spacings of quantum systems whose classical counterpart is chaotic agrees with the random matrix theory (RMT) predictions~\cite{mehta_91}. As a result, RMT statistics of level spacings has become a defining property of quantum chaotic systems. These ideas have been extended to lattice Hamiltonians in regimes that do not have a semiclassical limit~\cite{montambaux_poilblanc_93, hsu_dauriac_93, poilblanc_ziman_93, distasio_zotos_95, prosen_99, santos_04, rabson_narozhny_04, kolovsky_buchleitner_04, santos_rigol_10, rigol_santos_10, kollath_roux_10}.

One can divide lattice Hamiltonians into quadratic Hamiltonians and interacting ones. Quadratic Hamiltonians (our focus here) can be expressed in the diagonal form $\hat H = \sum_q \varepsilon_q \hat c_q^\dagger \hat c_q$, where $\{\varepsilon_q\}$ and $\{c_q^\dagger\}$ are the single-particle eigenenergies and the corresponding creation operators of the single-particle energy eigenstates, respectively. The many-body eigenstates of those Hamiltonians are products of single-particle eigenstates, $|q\rangle \equiv \hat c_q^\dagger |\emptyset\rangle$. As a consequence of this simple structure, quadratic models do not exhibit quantum chaos at the many-body level. Instead, RMT-like properties can be found in their single-particle sector. For example, the three-dimensional Anderson model below the localization transition~\cite{altshuler_shklovskii_86, altshuler_zharekeshev_88, shklovskii_shapiro_93, hofstetter_schreiber_93, sierant_delande_20} is known to exhibit single-particle level statistics and wave-function delocalization measures that agree with the RMT predictions. In what follows we refer to Hamiltonians that exhibit {\it single-particle} quantum chaos as quantum-chaotic quadratic Hamiltonians. They are to be contrasted to other quadratic Hamiltonians, e.g., translationally invariant ones, which do not exhibit single-particle quantum chaos. 

One of the motivations of our study is to show that it is possible to identify the underlying presence of {\it single-particle} quantum chaos in the {\it many-body} eigenstates of quadratic Hamiltonians. Recent studies of a class of quantum-chaotic quadratic models (SYK2 Hamiltonians) used the spectral form factor to demonstrate that single-particle quantum chaos manifests in the many-body spectrum as a residual repulsion between distant many-body energy levels~\cite{liao_vikram_20, winer_jian_20}.

Interacting (generic) Hamiltonians, on the other hand, cannot be reduced to bilinear forms in creation and annihilation operators. For those Hamiltonians, RMT-like properties are explored in the many-body context, e.g., by studying the spectral statistics of the many-body eigenenergies~\cite{montambaux_poilblanc_93, poilblanc_ziman_93, hsu_dauriac_93, distasio_zotos_95, prosen_99, santos_04, rabson_narozhny_04, kolovsky_buchleitner_04, santos_rigol_10, rigol_santos_10, kollath_roux_10} or the structure of their many-body eigenstates~\cite{karthik_sharma_07, santos_rigol_10, rigol_santos_10, kollath_roux_10, torresherrera_karp_16, mondaini_fratus_16, luitz_barlev_16, atas_bogomolny_17, beugeling_baecker_18, deutsch_10, vidmar_rigol_17, lu_grover_19, murthy_srednicki_19}. If properties of an interacting Hamiltonian agree with the RMT predictions, one says that the Hamiltonian is quantum chaotic. Such Hamiltonians are to be contrasted to integrable ones, for which exact solutions are available even in the many-body context in the presence of interactions.

Many studies have shown that entanglement measures are effective in identifying quantum chaos at the many-body level. They include, for example, bipartite~\cite{mejiamonasterio_beneti_05, santos_polkovnikov_12, deutsch13, beugeling_andreanov_15, yang_chamon_15, vidmar_rigol_17, garrison_grover_18, dymarsky_lashkari_18, nakagawa_watanabe_18, huang_19, lu_grover_19, murthy_srednicki_19, leblond_mallayya_19, wilming_goihl_19, morampudi_chandran_20, faiez_sefranek_20_a, faiez_sefranek_20_b, bhakuni_sharma_20, kaneko_iyoda_20, haque_mcclarty_20, miao_barthel_20, liu_chen_18, huang_gu_19, zhang_liu_20, alba_fagotti_09, moelter_barthel_14, lai_yang_15, nandy_sen_16, vidmar_hackl_17,  vidmar_hackl_18, hackl_vidmar_19, zhang_vidmar_18, jafarizadeh_rajabpour_19, Roy_2019, modak_nag_20, lydzba_rigol_20} and multipartite~\cite{karthik_sharma_07, brenes_pappalardi_20, lu_grover_20} entanglement measures on Hamiltonian eigenstates, and measures in the time domain that include the operator space entanglement entropy~\cite{zanardi_01, prosen_znidaric_07, znidaric_prosen_08, pizorn_prosen_09, alba_dubail_19} and entanglement entropy generation~\cite{modak_alba_20}. Here, we focus on the average bipartite (von Neumann and second R{\'e}nyi) entanglement entropy of Hamiltonian eigenstates. Recent studies, focusing on the von Neumann eigenstate entanglement entropy, showed that the average provides a useful tool to distinguish quantum-chaotic Hamiltonians~\cite{beugeling_andreanov_15, vidmar_rigol_17, garrison_grover_18, huang_19} from quadratic~\cite{vidmar_hackl_17, vidmar_hackl_18, hackl_vidmar_19} and interacting integrable~\cite{leblond_mallayya_19} ones.

Consider a bipartition of a lattice into two connected subsystems A (our subsystem of interest) and B. We define the subsystem fraction as $f=V_A/V$, the ratio between the volume $V_A$ (i.e., the number of lattice sites) of subsystem A and the total volume $V=V_A+V_B$. The reduced density matrix $\hat\rho_A$ of a many-body Hamiltonian eigenstate $|m\rangle$ in subsystem A is obtained by tracing out subsystem B, $\hat\rho_A = {\rm Tr}_B\{\hat \rho_m\}$, where $\hat \rho_m = |m \rangle \langle m|$. The von Neumann entanglement entropy of $|m\rangle$ is then defined as
\begin{equation}
S_{m} = - {\rm Tr} \{ \hat \rho_A \ln \hat\rho_A \}\,,
\end{equation}
and the corresponding second R{\'e}nyi entropy as
\begin{equation}
S_m^{(2)} = - \ln[{\rm Tr}\{ \hat \rho_A^2\}]\,.
\end{equation}
Throughout this work, we focus on systems with two states per lattice site (e.g., spinless fermions or spin-1/2 systems), and we are interested in the average (over all eigenstates) entanglement entropies
\begin{equation}\label{eq:aeee}
{\bar S} = 2^{-V} \sum_{m=1}^{2^V} S_m\,,
\end{equation}
and
\begin{equation}\label{eq:aree}
{\bar S}^{(2)} = 2^{-V} \sum_{m=1}^{2^V} S_m^{(2)}\,.
\end{equation}

\begin{figure}[!t]
\includegraphics[width=0.9\columnwidth]{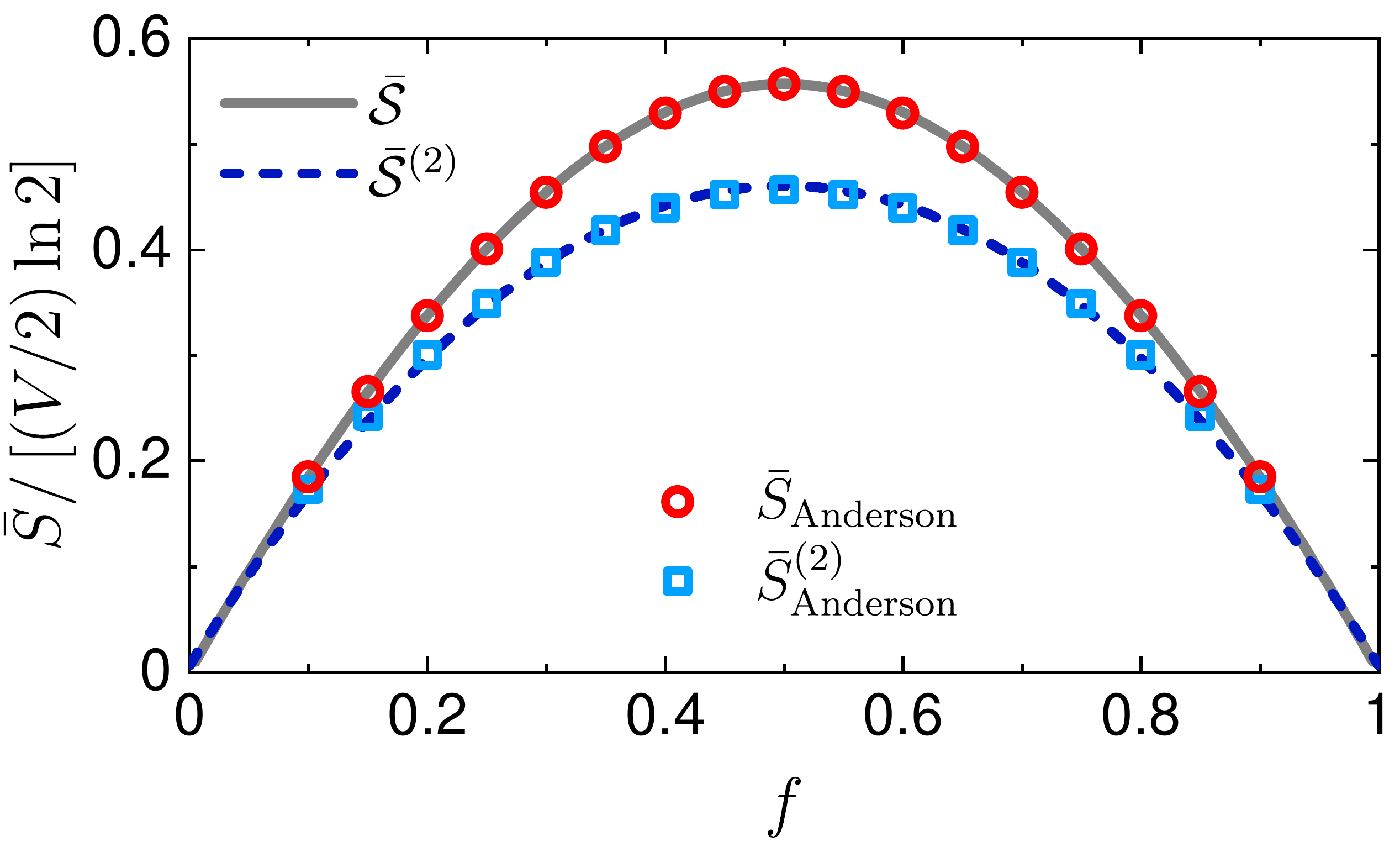}
\vspace{-0.1cm}
\caption{The average von Neumann $\bar{S}$ and second R{\'e}nyi $\bar{S}^{(2)}$ entanglement entropies as functions of a subsystem fraction $f$. The solid and dashed lines show the analytical predictions from Eqs.~$\left(\ref{eqAS}\right)$ and $\left(\ref{eqS}\right)$, respectively. Circles and squares show the numerical results for $\bar{S}$ and $\bar{S}^{(2)}$, respectively, in many-body eigenstates of the 3D Anderson model~(\ref{eqAn}) with disorder $W=1$ and $V=20^3=8000$ sites. The numerical results were obtained after averaging over $100$ randomly selected many-body eigenstates in each of 10 disorder realizations.}
\label{fig1}
\end{figure}

For the Hamiltonians studied here (see Sec.~\ref{sec:Hams}), the results for the averages coincide with the typical entanglement entropies of many-body energy eigenstates, so we use both terms interchangeably. We also note that for all model parameters under investigation, the average entanglement entropies exhibit a volume law scaling, i.e, the leading terms in $\bar S$ and $\bar S^{(2)}$ are proportional to the subsystem volume $V_A$.

One of the main results from previous studies of the average von Neumann entanglement entropy is that $\bar S$ exhibits a leading (volume-law) term of the form: (i)
\begin{equation} \label{def_Smax}
\bar S = V_A \ln 2
\end{equation}
for quantum-chaotic Hamiltonians~\cite{vidmar_rigol_17, garrison_grover_18, dymarsky_lashkari_18, huang_19, murthy_srednicki_19}, and (ii)
\begin{equation} \label{def_Squad}
\bar S = c_0(f) V_A \ln 2
\end{equation}
for quadratic Hamiltonians (in the absence of real-space localization)~\cite{storms_singh_14, vidmar_hackl_17, liu_chen_18, vidmar_hackl_18, hackl_vidmar_19} and for the translationally invariant (integrable) spin-1/2 XXZ chain~\cite{leblond_mallayya_19}. In Eq.~(\ref{def_Squad}), the volume-law coefficient $c_0(f=0) = 1$ and $c_0(f>0) < 1$. In other words, the average in Eq.~(\ref{def_Squad}) is not maximal if the subsystem fraction $f$ does not vanish in the thermodynamic limit.

Equation~(\ref{def_Smax}) tells us that the leading (in $V_A$) term in the entanglement entropy of typical eigenstates of quantum-chaotic Hamiltonians is the same as that for the thermodynamic entropy at the corresponding (``infinite-temperature'') energy~\cite{deutsch_10, deutsch13, vidmar_rigol_17, garrison_grover_18, dymarsky_lashkari_18}. Equation~(\ref{def_Smax}) also coincides with the leading term for the entanglement entropy averaged over random pure states in the Hilbert space obtained by Page~\cite{page_93}. 

In relation to Eq.~\eqref{def_Squad}, in Ref.~\cite{lydzba_rigol_20} we provided a closed-form expression for $c_0(f)$ obtained in the context of random quadratic Hamiltonians. For such Hamiltonians, the average von Neumann entanglement entropy of the eigenstates (for $f \leq 1/2$) reads
\begin{equation} \label{eqAS}
\bar{\mathcal{S}} = \left(1-\frac{1+f^{-1}\left(1-f\right) \ln\left(1-f\right)}{\ln 2}\right)V_A \ln 2 \,,
\end{equation}
see the solid line in Fig.~\ref{fig1}. (The results for $f>1/2$ are obtained replacing $V_A \to V - V_A$ and $f \to 1-f$.) To derive Eq.~\eqref{eqAS}, we assumed that the coefficients of the single-particle eigenstates are random and normally distributed in the position basis. Hence, one expects Eq.~(\ref{eqAS}) to be the analog of Page's result~\cite{page_93} for quadratic systems. The correctness of Eq.~(\ref{eqAS}) was checked numerically for many-body eigenstates of the SYK2 Hamiltonian in the Dirac fermion formulation and of power-law random banded matrices in the delocalized regime~\cite{lydzba_rigol_20}.

Here we conjecture that Eq.~(\ref{eqAS}) applies to quantum-chaotic quadratic Hamiltonians, which means that it allows one to identify the presence of single-particle quantum chaos in the many-body eigenstates of quadratic Hamiltonians. We test this conjecture for several quadratic Hamiltonians, ranging from those with nonlocal operators to those with only local operators. The latter include the well-known 3D Anderson model in the presence of weak disorder, see the circles in Fig.~\ref{fig1}. Going beyond the von Neumann entanglement entropy, we derive an analytic expression for the second R{\'e}nyi entanglement entropy (see the dashed line in Fig.~\ref{fig1}), and show numerical results (see the squares in Fig.~\ref{fig1} for 3D Anderson model results) that support the expectation that it is also universal for quantum-chaotic quadratic Hamiltonians. Note that the second R{\'e}nyi entanglement entropy can be viewed as a purity of a subsystem density matrix of a quantum state, and has been measured in experiments with ultracold atoms on optical lattices~\cite{islam_ma_15, kaufman_tai_16}.

The presentation is organized as follows. In Sec.~\ref{sec:Hams}, we introduce the quantum-chaotic quadratic Hamiltonians under investigation, and discuss the contrast between the single-particle and many-body results for the level spacing statistics when there is single-particle quantum chaos. We then study numerically, in Sec.~\ref{sec:vN}, the validity of Eq.~(\ref{eqAS}) in the many-body eigenstates of the models under investigation. In Sec.~\ref{sec:Renyi}, we derive an analytical expression for the second R{\'e}nyi entanglement entropy, which we then test numerically. We conclude with a summary and discussion in Sec.~\ref{sec:conclusions}.

\section{Quantum-chaotic quadratic Hamiltonians} \label{sec:Hams}

Quantum-chaotic quadratic Hamiltonians can be divided into two classes: local Hamiltonians, i.e., Hamiltonians $\hat H$ that are extensive sums of local operators [operators that have support on ${\cal O}(1)$ consecutive lattice sites], and nonlocal ones. Below we introduce those two classes separately.

\subsection{Nonlocal Hamiltonians} \label{subsec:nonlocal}

We consider the quadratic Sachdev-Ye-Kitaev model in the Dirac fermion formulation (in short, the Dirac SYK2 model),
\begin{equation} \label{dsyk2}
\hat {H}_\text{DSYK2} = \sum_{i,j=1}^{V} A_{ij} \hat f_i^\dagger \hat f^{}_j,
\end{equation}
as well as in the Majorana fermion formulation (the Majorana SYK2 model),
\begin{equation} \label{msyk2}
\hat {H}_{\text{MSYK2}} = \sum_{i,j=1}^{2V} \mathrm{i} \Lambda_{ij} \hat\chi_i \hat\chi_j,
\end{equation}
where $\hat f_i$ ($\hat f_i^\dagger$) is the Dirac fermion annihilation (creation) operator at site $i$, $V$ is the number of lattice sites for Dirac fermions, and $\hat\chi_i$ is the Majorana fermion operator. We assume that the neighboring Majorana fermions are paired, $\hat f_i = \hat\chi_{2i-1}+\mathrm{i} \hat\chi_{2i}$ and $\hat f_i^\dagger = \hat\chi_{2i-1}-\mathrm{i} \hat\chi_{2i}$~\cite{SYK_Kitaev_2001, Leijnse_2012}. The matrix ${\bf A}$ in Eq.~(\ref{dsyk2}) is a complex Hermitian matrix drawn from the Gaussian unitary ensemble, i.e., its diagonal elements are real numbers with zero mean and $2/V$ variance, while the off-diagonal elements are complex numbers with real and imaginary parts having zero mean and $1/V$ variance. The matrix ${\bf \Lambda}$ in Eq.~(\ref{msyk2}) is real and antisymmetric with normally distributed entries having zero mean and $\left(1+\delta_{ij}\right)/V$ variance~\cite{Sachdev_1993, liu_chen_18}.

We also consider the general quadratic (GQ) model
\begin{equation}
\label{eqGQ}
\hat H_\text{GQ}=\sum_{i,j=1}^{V} A_{ij} \hat f_i^\dagger \hat f_j + \sum_{i,j=1}^{V} \left( B_{ij} \hat f_i^\dagger \hat f_j^\dagger + B_{ij}^{*} \hat f^{}_j \hat f^{}_i \right)\,,
\end{equation}
for which the matrices ${\bf A}$ and ${\bf B}$ are complex Hermitian and complex antisymmetric, respectively. Their diagonal elements are normally distributed real numbers with zero mean and $2/V$ variance, while their off-diagonal elements are complex numbers with normally distributed real and imaginary parts with zero mean and $1/V$ variance. The Hamiltonian in Eq.~(\ref{eqGQ}) breaks the particle-number conservation present in Eq.~\eqref{dsyk2}. Note that the Majorana SYK2 model also breaks the particle-number conservation, i.e., the Hamiltonian in Eq.~(\ref{msyk2}) can be presented in the same form as the Hamiltonian in Eq.~$(\ref{eqGQ})$ when written in terms of Dirac fermions. However, the matrices ${\bf A}$ and ${\bf B}$ are related and fully determined by ${\bf \Lambda}$ in that case.

\subsection{Local Hamiltonians} \label{subsec:local}

We also study the Anderson model on a cubic lattice with $V$ sites~\cite{Anderson_1958},
\begin{equation} \label{eqAn}
\hat H_\text{A}=-\sum_{\langle i,j\rangle} \hat f_i^\dagger \hat f^{}_j + \frac{W}{2}\sum_{i=1}^V \epsilon_i \hat f_i^\dagger \hat f^{}_i \,,
\end{equation}
which is a local Hamiltonian with nearest-neighbor hopping and onsite disorder. The latter is described by independent uniformly distributed random numbers $\epsilon_i\in [-1,1]$, so that $W$ is the width of the disorder distribution. The indices in the sums in Eq.~(\ref{eqAn}) are defined as $i=x+\left(y-1\right)L+\left(z-1\right)L^2$ where $\left(x,y,z\right)$ are the Cartesian coordinates of a lattice site, each belonging to the set $[1,\ldots,L]$ with the linear size $L=V^{1/3}$. In the first sum in Eq.~(\ref{eqAn}), $\langle i,j\rangle$ denotes nearest neighbor sites $i$ and $j$. We use periodic boundary conditions so the Hamiltonian is translationally invariant at $W=0$.

The 3D Anderson model in Eq.~(\ref{eqAn}) exhibits a delocalization-localization transition upon increasing $W$. For single-particle eigenstates at the center of the energy spectrum, the critical value of $W$ is $W^*\approx 16.5$~\cite{Schubert_2005, slevin_ohtsuki_14, Slevin_2018}. In this work we focus on the delocalized regime $W < W^*$, and, in the context of the average eigenstate entanglement entropy, in the regime in which the overwhelming majority of the single-particle eigenstates are delocalized for the system sizes studied.

We note that the 3D Anderson model conserves the number of particles. We also study an extended 3D Anderson model ($\hat H_{\text{EA}}$) that contains local terms that break particle-number conservation
\begin{equation}\label{eqH}
\hat H_{\text{EA}} =  \hat H_\text{A} + \sum_{\langle i,j\rangle} \Delta_{ij}\left( \hat f_i^\dagger \hat f_j^\dagger + \hat f^{}_j \hat f^{}_i \right), 
\end{equation}
where $\Delta_{ij} = \text{sign}(i-j)\Delta$. When $W=0$, the Hamiltonian in Eq.~(\ref{eqH}) is somewhat related to the Hamiltonian of $p$-wave superconductors with a pairing field $\Delta$~\cite{kallin_berlinsky_16}. We therefore refer to $\Delta$ as the strength of the pairing field.

Quadratic Hamiltonians that break particle-number conservation can be diagonalized using a Bogoliubov transformation, $\hat c_q = \sum_j \alpha_{qj} \hat f_j+\beta_{qj} \hat f_{j}^\dagger$. The inverse transformation is $\hat f_j=\sum_{q} v_{jq} \hat c_q+\theta_{jq} \hat c_q^\dagger$, where $\alpha_{qj} = v_{jq}^{*}$ and $\beta_{qj}=\theta_{jq}$. Hence, the ``rotated'' Hamiltonians commute with the quasiparticle-number operators $\hat c_q^\dagger \hat c^{}_q$~\cite{Bogoliubov_1958, Fetter_2003}. For particle-number conserving quadratic Hamiltonians, $\hat c_q = \sum_j \alpha_{qj} \hat f_j$.  

\subsection{Level spacing statistics} \label{sec:stat}

The statistics of the spacings $\delta_i = E_i-E_{i-1}$ between nearest neighbor energy levels $E_{i-1}$ and $E_i$ is commonly used to identify quantum chaos. It is convenient to introduce the ratio $\tilde{r}_i = {\rm min} \{\delta_i,\delta_{i+1}\} / {\rm max} \{\delta_i,\delta_{i+1}\}$, such that $\tilde r_i \in [0,1]$, which allows one to study statistical properties without the need of carrying out a spectral unfolding~\cite{oganesyan_huse_07}. Within the Gaussian orthogonal ensemble (GOE) of RMT, the distribution of ratios $\tilde r_i$ was obtained analytically for $3\times 3$ matrices~\cite{atas_bogomolny_13}
\begin{equation} \label{def_Pr}
P\left(\tilde{r}\right)=\frac{27}{4}\frac{\left(\tilde{r}+\tilde{r}^2\right)}
{\left(1+\tilde{r}+\tilde{r}^2\right)^\frac{5}{2}} \,.
\end{equation}
Numerical studies of the energy level statistics in quantum-chaotic Hamiltonians have shown that Eq.~\eqref{def_Pr} closely follows the exact distribution of ratios $\tilde r_i$ in large systems~\cite{mondaini_fratus_16, jansen_stolpp_19}.

\begin{figure}[!t]
\includegraphics[width=0.8\columnwidth]{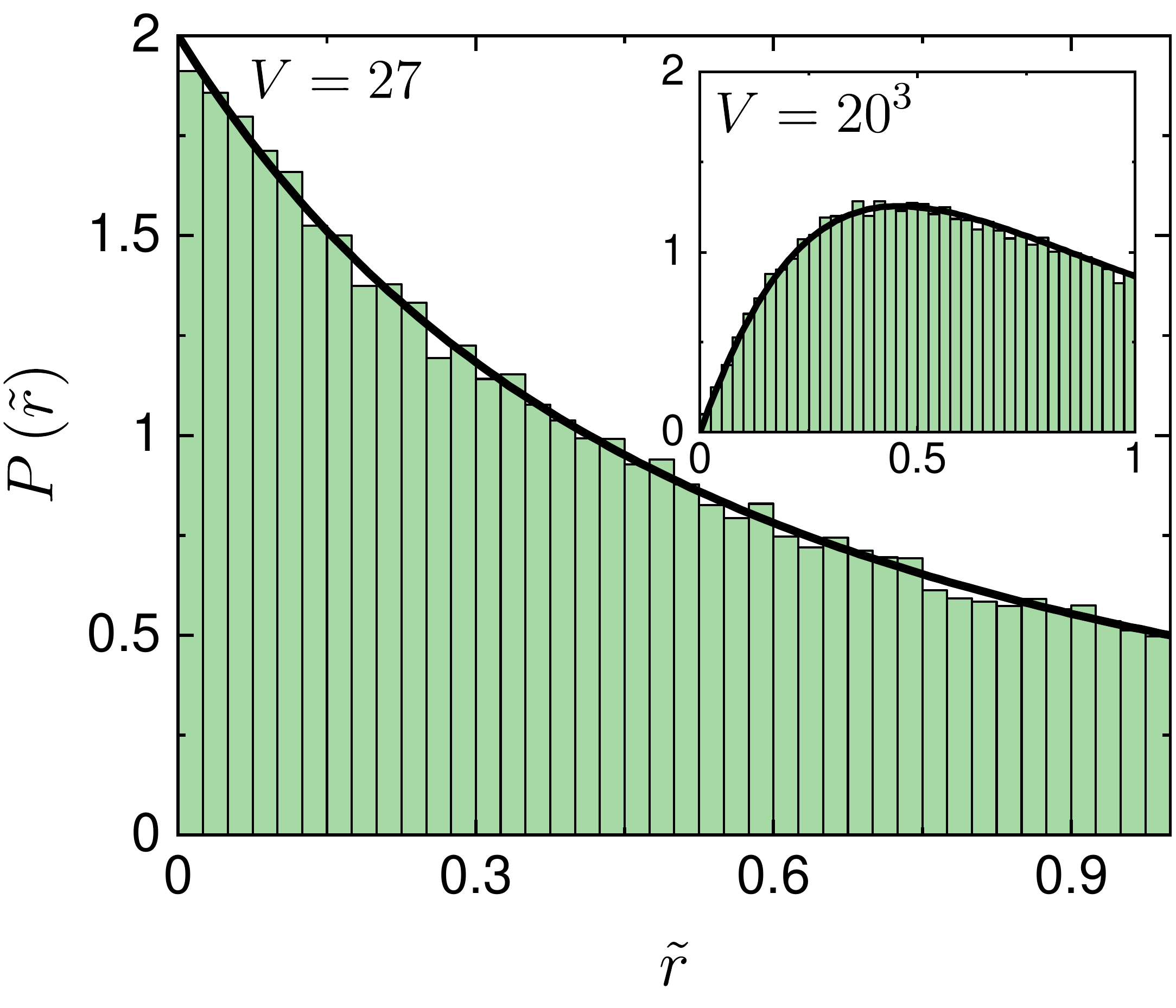}
\vspace{-0.1cm}
\caption{The distribution $P(\tilde r)$ of level spacing ratios in the 3D Anderson model~(\ref{eqAn}) for $W=1$. Main panel: $P(\tilde r)$ for the many-body level spacings on $V=27$ sites. The solid line is the distribution for uncorrelated random energy levels~(\ref{def_Pr_poisson}). The numerical results were obtained using $500$ (out of the $2^{27}$) many-body eigenstates about the mean energy of the spectrum and averaging over 100 disorder realizations. Inset: $P(\tilde r)$ for the single-particle level spacings on $V=20^3=8000$ sites. The solid line is the GOE result for $3\times3$ matrices~(\ref{def_Pr}). The numerical results were obtained using $500$ (out of $20^3$) single-particle eigenstates about the mean energy of the spectrum, and averaging over 100 disorder realizations.}
\label{fig2}
\end{figure}

$P\left(\tilde{r}\right)$, when used for the single-particle energy spectrum, can also be used to detect single-particle quantum chaos. The nonlocal SYK2-like models introduced in Sec.~\ref{subsec:nonlocal} comply with the RMT predictions by construction. It is also well known that the level-spacing statistics of the 3D Anderson model~(\ref{eqAn}) complies with the RMT predictions in the delocalized regime~\cite{Evangelou_1991, Mirlin_2000, Evers_2008}. In the inset in Fig.~\ref{fig2} we show the agreement between $P(\tilde r)$ in the 3D Anderson model and the predictions of Eq.~(\ref{def_Pr}). Moreover, in the inset of Fig.~\ref{fig5}(b), we show that the distribution remains unchanged in the extended 3D Anderson model in Eq.~(\ref{eqH}).

The main panel in Fig.~\ref{fig2} shows that, in contrast to the single-particle results in the inset, $P\left(\tilde{r}\right)$ for the many-body eigenstates of the 3D Anderson model exhibits no level repulsion (as expected). The numerical results for $P\left(\tilde{r}\right)$ in the main panel agree with the analytic predictions for uncorrelated random energy levels~\cite{atas_bogomolny_13}
\begin{equation} \label{def_Pr_poisson}
P\left(\tilde{r}\right)= \frac{2}{(1+\tilde r)^2} \,.
\end{equation}
Therefore, at the energy scale that corresponds to the mean level spacing of the many-body spectrum, as shown in Fig.~\ref{fig2}, the many-body energy levels of quantum-chaotic quadratic Hamiltonians behave as uncorrelated random numbers. This is exactly the way in which the energy levels of integrable interacting Hamiltonians behave~\cite{hsu_dauriac_93, poilblanc_ziman_93, santos_rigol_10}. Interestingly, two recent works~\cite{liao_vikram_20, winer_jian_20} showed that the many-body spectral statistics of quantum-chaotic quadratic Hamiltonians exhibit deviations from the predictions for uncorrelated random energy levels at energy scales comparable or larger than the single-particle level spacing.

\section{von Neumann entanglement entropy} \label{sec:vN}

\subsection{Theoretical considerations}

We first describe the main steps needed to calculate the von Neumann entanglement entropy of the many-body eigenkets $\{|m\rangle\}$ of the Hamiltonians introduced in Sec.~\ref{sec:Hams}. We denote the single-particle energy eigenkets as $\{ |q\rangle = \hat c_q^\dagger |\emptyset\rangle;\, q = 1, ..., V \}$. The many-body eigenkets are then constructed as $\{|m\rangle = \prod_{\{q_l\}_m} \hat c_{q_l}^\dagger |\emptyset\rangle;\, m = 1, ..., 2^V\}$, where $\{ q_l\}_m$ represent the $m$th set of occupied single-particle energy eigenkets.

If the number of particles is conserved, all many-body correlations for the eigenket $\ket{m}$ can be computed (via Wick's theorem) using the $V\times V$ generalized one-body correlation matrix~\cite{chung_peschel_01, peschel_03, peschel_eisler_09, vidmar_hackl_17, hackl_bianchi_20}
\begin{equation} \label{rhom}
(\mathcal{J}_m)_{ij} = \bra{m}\hat f_i^\dagger \hat f_j - \hat f_j \hat f_i^\dagger \ket{m} = 2(\rho_m)_{ij} - \delta_{ij} \,,
\end{equation}
where $\rho_m$ is the one-body correlation matrix of $\ket{m}$. The matrix $\mathcal{J}_m$ can be written as
\begin{equation} \label{Jm_ij}
(\mathcal{J}_m)_{ij} = \sum_{q=1}^{V} \bra{m}2 \hat c^\dagger_q \hat c_q-1\ket{m} v_{iq}^* v_{jq} \,,
\end{equation}
where $(2 \hat c^\dagger_q \hat c^{}_q-1)|m\rangle = 1$ $(-1)$ for an occupied (empty) single-particle energy eigenket $|q\rangle$ in the many-body energy eigenket $|m\rangle$.

To calculate the von Neumann entanglement entropy of eigenket $|m\rangle$, we bipartition the system into a connected subsystem A with $V_A$ lattice sites and an environment B with $V-V_A$ lattice sites, and restrict $\mathcal{J}_m$ in Eq.~(\ref{Jm_ij}) to the entries with $i,j$ from the subsystem A. Then, the von Neumann entanglement entropy can be obtained as~\cite{Bengtsson_2006, peschel_eisler_09}
\begin{equation}
S_m = -\sum_{i=1}^{V_A} \left(\frac{1+\lambda_i}{2} \ln\left[\frac{1+\lambda_i}{2}\right] + \frac{1-\lambda_i}{2} \ln\left[\frac{1-\lambda_i}{2}\right] \right) \,,
\end{equation}
where $\{\lambda_i\}$ are the eigenvalues of the restricted $\mathcal{J}_m$.

If the number of particles is not conserved, all many-body correlations for eigenket $\ket{m}$ can be computed using the $2V\times 2V$ generalized one-body correlation matrix~\cite{vidmar_hackl_17, hackl_bianchi_20}
\begin{equation} \label{def_Jm}
J_m = \left[ \begin{split}
 &\mathcal{J}_m &\mathcal{K}_m \\ 
-&{\mathcal{K}_m}^* &-{\mathcal{J}_m}^*
\end{split} \right] \,,
\end{equation}
where the matrix elements of $\mathcal{J}_m$ [defined in Eq.~\eqref{rhom}] can be written as
\begin{equation}
(\mathcal{J}_m)_{ij} = \sum_{q=1}^{V} \bra{m}2 \hat c^\dagger_q \hat c_q-1\ket{m} \left(v_{iq}^*v_{jq} - \theta_{iq}^* \theta_{jq}\right),    
\end{equation}
and~\cite{cheong_henley_04, peschel_03}
\begin{equation}
(\mathcal{K}_m)_{ij} = \bra{m} \hat f_i^\dagger \hat f_j^\dagger - \hat f_j^\dagger \hat f_i^\dagger \ket{m} \,,
\end{equation}
which can be written in terms of the parameters of the Bogoliubov transformation as
\begin{equation}
(\mathcal{K}_m)_{ij} = \sum_{q=1}^{V} \bra{m}2 \hat c^\dagger_q \hat c_q-1\ket{m}\left(v_{iq}^*\theta_{jq}^*-\theta_{iq}^* v_{jq}^*\right). 
\end{equation}

In this case, the von Neumann entanglement entropy can be calculated from $J_m$ restricted to the entries of $\mathcal{J}_m$ and $\mathcal{K}_m$ with $i,j$ in subsystem A,
\begin{equation}
S_m = -\sum_{i=1}^{2V_A} \frac{1+\tilde{\lambda}_i}{2} \ln\left[\frac{1+\tilde{\lambda}_i}{2}\right] \,,
\end{equation}
where $\{\tilde{\lambda}_i \}$ are eigenvalues of the restricted $J_m$.

\subsection{Numerical results}

\begin{figure}[!b]
\includegraphics[width=0.9\columnwidth]{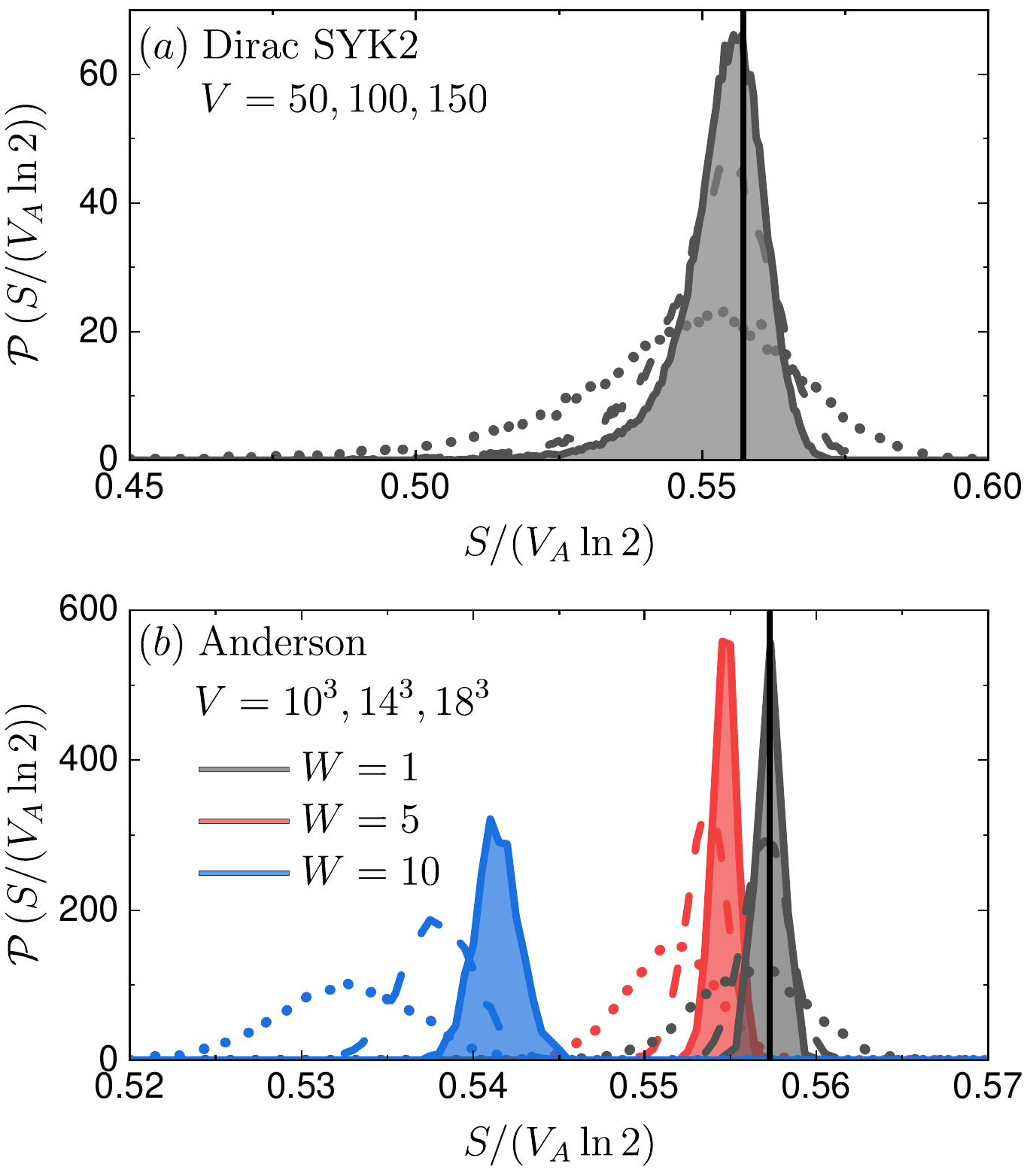}
\vspace{-0.1cm}
\caption{Distributions of the von Neumann eigenstate entanglement entropies at $f=1/2$ for (a) the Dirac SYK2 model~(\ref{dsyk2}), and (b) the 3D Anderson model~(\ref{eqAn}) with $W=1$, 5, and 10. The vertical lines mark the analytical prediction [Eq.~(\ref{eqAS})]. We show results for lattices with: (a) $V = 50$, 100, and 150, and (b) $V=10^3,\,14^3,$ and $18^3$, using dotted, dashed, and solid lines, respectively. The results in (a) were obtained using at least $10^5$ randomly selected eigenstates, in each of the $500$ Hamiltonian realizations over which we average. The results in (b) were obtained for $10^4$, $10^3$, and $10^2$ randomly selected eigenstates in each of the 100, 10, and 5 Hamiltonian realizations for $V=10^3,\,14^3$, and $18^3$, respectively.}\label{fig3}
\end{figure}

In Fig.~\ref{fig3}, we show the normalized distribution of eigenstate entanglement entropies $S_m/[V_A \ln 2]$ for the Dirac SYK2 model~(\ref{dsyk2}) [Fig.~\ref{fig3}(a)] and the 3D Anderson model~(\ref{eqAn}) [Fig.~\ref{fig3}(b)]. In both cases, the distribution narrows upon increasing the system size. This is consistent with results reported in Ref.~\cite{lydzba_rigol_20}, which showed that the variance $\sigma^2= 2^{-V} \sum_{m=1}^{2^V} \left(S_m - \bar{S}\right)^2/(V_A\ln 2)^2$ of the distribution [with $\bar{S}$ as defined in Eq.~\eqref{eq:aeee}] in the Dirac SYK2 model vanishes as $V\rightarrow\infty$. Therefore, the average entanglement entropy over all eigenstates $\bar{S}$, the quantity on which we focus in the remainder of this paper, coincides with the typical eigenstate entanglement entropy as $V\rightarrow\infty$.

In our numerical calculations we approximate the average over all eigenstates by an average over a set of randomly selected eigenstates with equal probability throughout the spectrum. Those averages are then further averaged over different realizations of the random Hamiltonians under consideration.
In Appendix~\ref{sec:app1} we show, studying the realization-to-realization fluctuations for the 3D Anderson model at $W=1$, that the average entanglement entropy over eigenstates of a single Hamiltonian provides an accurate estimate for the entanglement entropy obtained after averaging over disorder realizations.

\begin{figure}[!t]
\includegraphics[width=0.95\columnwidth]{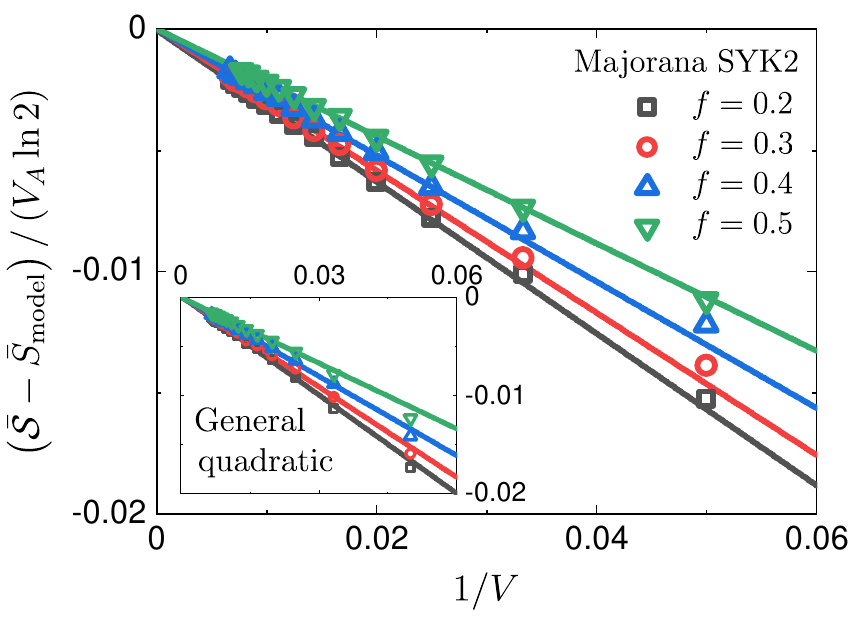}
\vspace{-0.1cm}
\caption{Average von Neumann entanglement entropies $\bar S_{\rm model}$ of eigenstates of: (main panel) the Majorana SYK2 model~(\ref{msyk2}) and (inset) the general quadratic model~(\ref{eqGQ}). The averages were computed over at least $10^5$ randomly selected many-body eigenstates, and from $100$ to $500$ Hamiltonian realizations. Results for $\bar S_{\rm model}$ are shown for different subsystem fractions $f$, as indicated in the legend, and subtracted from the analytical prediction $\bar{\cal S}$ in Eq.~(\ref{eqAS}). Solid lines show the results of two-parameter linear fits to $a_0+a_1/V$ using the data for $V\geq 70$. In all cases we get $a_0\leq 5\cdot 10^{-5}$.}\label{fig4}
\end{figure}

As shown in Ref.~\cite{lydzba_rigol_20}, the leading volume-law term in the average entanglement entropy $\bar S$ of the Dirac SYK2 model~(\ref{dsyk2}) agrees in the thermodynamic limit with the analytical expression $\bar{\mathcal{S}}$ from Eq.~(\ref{eqAS}). It was pointed out in Ref.~\cite{liu_chen_18} that, because of the particle-hole symmetry, $\bar S$ in the Majorana SYK2 model~(\ref{msyk2}) exhibits the same volume-law contribution. This is numerically confirmed in the main panel of Fig.~\ref{fig4}, where we show that the difference between the numerical results in finite systems and the analytic prediction appear to vanish with increasing system size. The results in the inset in Fig.~\ref{fig4}, for the general quadratic model in Eq.~(\ref{eqGQ}), show that leading volume-law term in the average eigenstate entanglement entropy is actually insensitive to particle-number conservation. For all results in Fig.~\ref{fig4}, the extrapolated thermodynamic limit value of the volume-law coefficient agrees with the predictions in Eq.~(\ref{eqAS}) to within $\lim_{V,V_A \to \infty}|\bar{\mathcal{S}} - \bar{S}_{\rm model}|/(V_A \ln 2) \leq 5\cdot 10^{-5}$.

We stress that the difference between the particle-number conserving and nonconserving models can be detected already in the first subleading term of the average entanglement entropy. For the nonlocal models introduced in Sec.~\ref{subsec:nonlocal}, the dominant subleading term at $f>0$ is a constant. This constant is positive for the particle nonconserving models such as the Majorana SYK2 model and the general quadratic model (see Fig.~\ref{fig4}), and negative for the particle conserving Dirac SYK2 model (see Fig.~1 of Ref.~\cite{lydzba_rigol_20}).

\begin{figure}[!t]
\includegraphics[width=0.95\columnwidth]{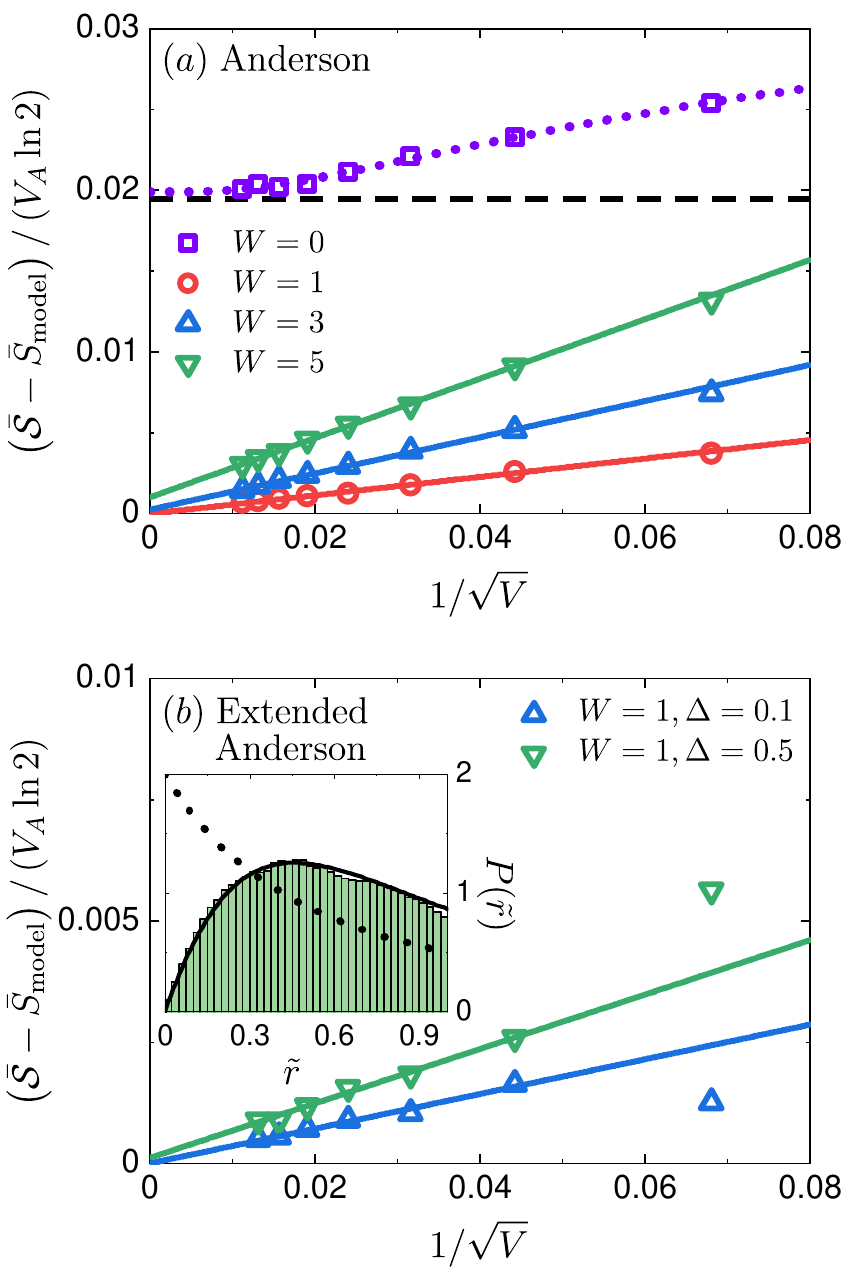}
\vspace{-0.1cm}
\caption{Average von Neumann entanglement entropies $\bar S_{\rm model}$ of eigenstates of: (a) the 3D Anderson model~(\ref{eqAn}) and (b) the extended 3D Anderson model~(\ref{eqH}), at subsystem fraction $f=1/2$. The averages were computed over $10^2$ to $10^4$ randomly selected many-body eigenstates, in from $5$ to $500$ Hamiltonian realizations. The solid lines show the results of two-parameter fits to $a_0+a_1/\sqrt{V}$ using the data for $V \geq 512$. We get $a_0\leq 5\cdot 10^{-4}$ for $W = 1$ and 3, and $a_0 = 9\cdot 10^{-4}$ for $W = 5$. The dotted line in (a) corresponds to a three-parameter fit $a_0+a_1 e^{-a_2 L}$ (with $L=V^{1/3}$) to the data at $W=0$ with $V \geq 512$~\cite{thermodynamic2}.
The horizontal dashed line marks the thermodynamic limit result obtained numerically for translationally-invariant noninteracting fermions in one dimension $\bar{S}_{W=0}/(V_A\ln 2) \approx 0.5378$~\cite{vidmar_hackl_17}. Inset in (b): The distribution of ratios of level spacings $P\left(\tilde{r}\right)$ at $W=1, \Delta=0.5$, and $V=8000$, for $500$ single-particle states around the mean energy of the single-particle spectrum. The solid and dotted lines show the results of Eqs.~(\ref{def_Pr}) and~(\ref{def_Pr_poisson}), respectively.}\label{fig5}
\end{figure}

We next show that the result in Eq.~(\ref{eqAS}) accurately describes the volume-law contribution to the average entanglement entropies $\bar S$ of quantum-chaotic quadratic Hamiltonians that are sums of local operators in position space (local quantum-chaotic quadratic Hamiltonians). For this, we focus on the 3D Anderson model at $f=1/2$ (see Fig.~\ref{fig1} for results as a function of $f$). In Fig.~\ref{fig5}(a) we show results for the traditional (particle-number conserving) 3D Anderson model~(\ref{eqAn}) in the delocalized regime, and in Fig.~\ref{fig5}(b) we show results for the extended (particle-number nonconserving) Anderson model~(\ref{eqH}). In both cases one can see that the difference between the numerical results in finite systems and the analytic prediction decreases with increasing system size. For all results in Fig.~\ref{fig5}, we find that $\lim_{V,V_A \to \infty} |\bar{\mathcal{S}} - \bar{S}_{\rm model}|/(V_A \ln 2) \leq 5\cdot 10^{-4}$ for $W \leq 3$.

We note that there is a significant difference between the local and nonlocal models in terms of the first subleading correction to the leading volume-law term at $f>0$. While the dominant subleading correction for nonlocal models is a constant (cf.~Fig.~\ref{fig4}), it grows subextensively for the local models (cf.~Fig.~\ref{fig5}). The fits in Fig.~\ref{fig5} for the 3D Anderson models at $f=1/2$ suggest that $\bar S = \bar{\cal S} + {\cal O}(\sqrt{V})$. A similar subleading contribution was observed for the average entanglement entropy over eigenstates of interacting integrable models~\cite{leblond_mallayya_19}. 
With regard to the results for the particle-number conserving model in Fig.~\ref{fig5}(a), we note that for $W=0$ it corresponds to translationally invariant noninteracting fermions in a cubic lattice, for which the single-particle level statistics does not obey the RMT predictions~\cite{cheng_mondaini_16, Torres_Herrera_2019}. The results in Fig.~\ref{fig5}(a) for $\bar S$ at $W=0$ show that $\lim_{V,V_A \to \infty} |\bar{\mathcal{S}} - \bar{S}_{\rm model}|/(V_A \ln 2) \neq 0$. Instead, $\bar S$ approaches the result for translationally-invariant noninteracting fermions in one dimension obtained in Ref.~\cite{vidmar_hackl_17} [shown as a horizontal dashed line in Fig.~\ref{fig5}(a)]. This provides further support to the conjecture put forward in Ref.~\cite{vidmar_hackl_18} that the average entanglement entropy is universal for all translationally invariant quadratic models. While in Ref.~\cite{hackl_vidmar_19} tight bounds were provided for that case, the corresponding close-form expression (if any) remains elusive.

A second important observation about the results in Fig.~\ref{fig5}(a) is that the numerically obtained $\lim_{V,V_A \to \infty} |\bar{\mathcal{S}} - \bar{S}_{\rm model}|/(V_A \ln 2)$ increases with increasing the magnitude of $W$. The small offset is larger for $W=5$ (of the order $10^{-3}$) when compared to results for $W\leq 3$. This is expected in the light of the mobility edge in the single-particle spectrum of the 3D Anderson model. Namely, one expects the differences between the exact averages over all eigenstates and the predictions from Eq.~\eqref{eqAS} to persist in the thermodynamic limit whenever there are localized states at the edges of the spectrum, and to decrease as the fraction of localized states decreases. Hence, we stress that our numerical results show that the closed-form expression in Eq.~\eqref{eqAS} provides an accurate estimate for the average eigenstate entanglement entropy in the 3D Anderson model at weak disorder (far below $W^*$). It is interesting to note that even at $W=10$, as one can conclude from the results shown in Fig.~\ref{fig3}(b), the average entanglement entropy is close to the result in Eq.~(\ref{eqAS}). The functional form of the difference between the results for finite $W<W^*$ in the thermodynamic limit and the predictions of Eq.~(\ref{eqAS}) is yet to be established.

\section{The second R{\'e}nyi entropy} \label{sec:Renyi}

\subsection{Theoretical considerations}

The numerical calculation of $S_m^{(2)}$ is similar to the one for the von Neumann entanglement entropy in Sec.~\ref{sec:vN}, i.e., one needs to construct an appropriate one-body correlation matrix and obtain the corresponding eigenvalues. For particle-number conserving systems~\cite{hackl_bianchi_20}
\begin{equation} \label{S2_m}
S_{m}^{(2)} = - \sum_{i=1}^{V_A}\ln\left[\alpha_i^2 + \left(1-\alpha_i\right)^2\right] \,,
\end{equation}
where $\left\{\alpha_i;\, i=1,...,V_A\right\}$ are the eigenvalues of the one-body correlation matrix $\rho_m$ from Eq.~(\ref{rhom}), restricted to entries that belong to subsystem A.

In order to compute ${\bar S}^{(2)}$ [defined in Eq.~\eqref{eq:aree}] analytically, we follow a different approach to the one in Ref.~\cite{lydzba_rigol_20}. We first note that in Ref.~\cite{liu_chen_18} it was pointed out (and tested numerically) that the restricted correlation matrix $\rho_m$~(\ref{rhom}) of a typical many-body eigenket $\ket{m}$ of the Dirac SYK2 model~(\ref{dsyk2}) belongs to the $\beta$-Jacobi ensemble with $\beta=2$. The distribution of its eigenvalues, for different subsystem fractions $f$, has the following form~\cite{liu_chen_18}:
\begin{equation} \label{eqF}
\mathcal{F}_f\left(\alpha\right) = \frac{1}{2\pi f} \frac{\sqrt{\alpha\left(1-\alpha\right) + f\left(1-f\right)-\frac{1}{4}}}{\alpha\left(1-\alpha\right)} \, 1_{\left[\alpha_{-},\alpha_{+}\right]} \,,
\end{equation}
where we assumed that the system is at half filling. $\mathcal{F}_f\left(\alpha\right)$ is nonzero for $\alpha \in \left[\alpha_{-},\alpha_{+}\right]$, where $\alpha_{\pm}=\frac{1}{2}\pm\sqrt{f\left(1-f\right)}$. Using Eq.~(\ref{eqF}), one can calculate the average second R{\'e}nyi entanglement entropy as $\bar{\mathcal{S}}^{(2)} = \int d\alpha\, \mathcal{F}_f\left(\alpha\right) S^{(2)}(\alpha)$ by replacing the sum in Eq.~(\ref{S2_m}) by the integral, $\sum_i \to V_A \int d\alpha {\cal F}_f(\alpha)$.

We write $\alpha=\left(\lambda+1\right)/2$, where $\lambda$ are the eigenvalues of the restricted one-body correlation matrix ${\cal J}$ in Eq.~(\ref{rhom}), to obtain
\begin{equation} \label{defS2_int}
\begin{split}
\bar{\mathcal{S}}^{(2)} = & -\frac{V_A}{\pi f} \int_{\lambda_{-}}^{\lambda_{+}}  \frac{\sqrt{\frac{1}{4}\left(1-\lambda^2\right) + f\left(1-f\right)-\frac{1}{4}}}{1-\lambda^2}\\ & \qquad \qquad \quad \times\ln\left[1-\frac{1}{2}\left(1-\lambda^2\right)\right] d\lambda \,,
\end{split}
\end{equation}
where $\lambda_{\pm}=\pm\sqrt{4f\left(1-f\right)}$. Next, we replace the logarithm with its power series expansion, which yields
\begin{equation} \label{defS2_series}
\begin{split}
\bar{\mathcal{S}}^{(2)} & = \frac{V_A}{\pi f} \sum_{k=1}^{\infty}\frac{2^{-k}}{k} \int_{\lambda_{-}}^{\lambda_{+}} \left(1-\lambda^2\right)^{k-1} \\
& \qquad \qquad \times \sqrt{\frac{1}{4}\left(1-\lambda^2\right) + f\left(1-f\right)-\frac{1}{4}}\;d\lambda \, .\\
\end{split}
\end{equation}
Each integral in the sum in Eq.~(\ref{defS2_series}) can be written in terms of a hypergeometric function,
\begin{equation}
\begin{split}
& \int_{\lambda_{-}}^{\lambda_{+}} \left(1-\lambda^2\right)^{k-1}\sqrt{\frac{1}{4}\left(1-\lambda^2\right) + f\left(1-f\right)-\frac{1}{4}}\;d\lambda \\ & = f\left(1-f\right) \pi\;  {}_2F_{1}\left[\frac{1}{2},\,1-k,\,2,\,4f\left(1-f\right)\right] \,, \\
\end{split}
\end{equation}
with $_2F_{1}\left(a,b,c,d\right) = \sum_{n=0}^{\infty} \frac{\left(a\right)_n \left(b\right)_n}{\left(c\right)_n} \frac{d^n}{n!}$, where $\left(a\right)_0=1$, and $\left(a\right)_n = a\left(a+1\right)...\left(a+n-1\right)$ for integer $n>0$ is known as a Pochhammer symbol. As a result, the average second R{\'e}nyi entropy can be expressed as an infinite series
\begin{equation}\label{eqS}
\bar{\mathcal{S}}^{(2)} =V_A\left(1-f\right) \sum_{k=1}^{\infty}\frac{2^{-k}}{k} \;{}_2F_{1}\left[\frac{1}{2},\,1-k,\,2,\,4f\left(1-f\right)\right].
\end{equation}
The hypergeometric function for $f<1/2$ and $k\geq 1$ is a polynomial of $4f\left(1-f\right)$ of degree $k-1$. The special point $f=1/2$ is given by the Chu-Vandermonde identity ${}_2F_{1} \left(1/2,1-k,2,1\right) = \frac{\left(3/2\right)_{1-k}} {\left(2\right)_{1-k}}$~\cite{DLMF}. Therefore, $\bar{\mathcal{S}}^{(2)}$ always converges to a finite value, as can be verified with the help of the ratio test.

Equation~(\ref{eqS}) reduces to simple closed-form expressions in two limiting cases. When $f\to 0$, the distribution in Eq.~(\ref{eqF}) becomes a delta function at $\alpha=1/2$ and hence the entanglement entropy is maximal $\bar{\mathcal{S}}^{(2)} \rightarrow V_A \ln 2$ (see also Ref.~\cite{vidmar_hackl_17}). When $f=1/2$, Eq.~(\ref{defS2_int}) becomes
\begin{equation} \label{S2_f12}
\bar{\mathcal{S}}^{(2)} = \left(2+2\frac{\ln\left(2-\sqrt{2}\right)}{\ln 2}\right) V_A \ln 2\approx 0.457 \, V_A \ln 2 \,.
\end{equation}

In what follows we test whether $\bar{\mathcal{S}}^{(2)}$~\eqref{eqS} is universal for the average second R{\'e}nyi entanglement entropy of eigenstates of quantum-chaotic quadratic Hamiltonians, in the same way that $\bar{\mathcal{S}}$~\eqref{eqAS} is for the von Neumann entanglement entropy. We compare $\bar{\mathcal{S}}^{(2)}$~\eqref{eqS} to the numerical evaluation of $\bar{S}^{(2)}$ as defined in Eq.~\eqref{eq:aree}. For particle-conserving models, $S_m^{(2)}$ is computed evaluating Eq.~(\ref{S2_m}).

We also conjecture that Eq.~\eqref{eqS} is valid for models the particle number is not conserved. In that case, $S_m^{(2)}$ is computed evaluating 
\begin{equation}\label{eq:s2f2}
S_{m}^{(2)}=-\frac{1}{2}\sum_{i=1}^{2V_A}\ln\left[ \frac{1 + \tilde{\lambda}_i^2}{2} \right] \,,
\end{equation}
where $\{\tilde{\lambda}_i \}$ are the eigenvalues of the restricted one-body correlation matrix $J_m$ from Eq.~(\ref{def_Jm}).

\subsection{Numerical results}

\begin{figure}[!b]
\includegraphics[width=0.95\columnwidth]{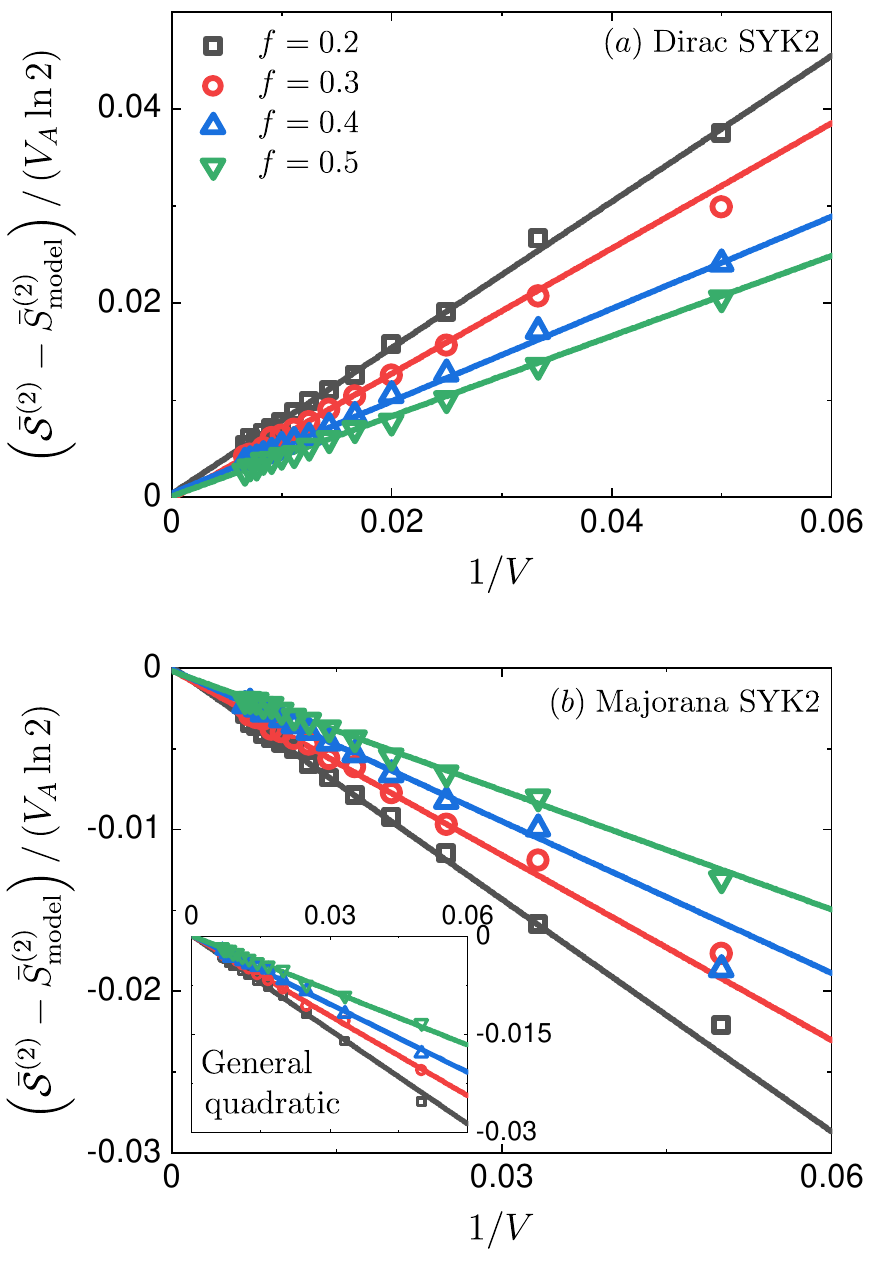}
\vspace{-0.1cm}
\caption{Average second R{\'e}nyi entanglement entropies $\bar S_{\rm model}^{(2)}$ of eigenstates of: (a) the Dirac SYK2 model~(\ref{dsyk2}), (b) the Majorana SYK2 model~(\ref{msyk2}), and inset in (b) the general quadratic model~(\ref{eqGQ}). The averages we computed as in Fig.~\ref{fig4}. Results for $\bar S_{\rm model}^{(2)}$ are shown for different subsystem fractions $f$, as indicated in the legend, and subtracted from the analytical prediction $\bar{\cal S}^{(2)}$ in Eq.~(\ref{eqS}). Solid lines are the results of two-parameter linear fits to $a_0+a_1/V$ using the data for $V\geq 70$. In all cases we get $a_0\leq 4\cdot 10^{-4}$.
}\label{fig6}
\end{figure}

In Fig.~\ref{fig6}, we compare $\bar{\mathcal{S}}^{(2)}$~\eqref{eqS} with numerical results for the average second R{\'e}nyi entanglement entropy $\bar S^{(2)}$ of eigenstates of the nonlocal quantum-chaotic quadratic Hamiltonians introduced in Sec.~\ref{subsec:nonlocal}. The comparison yields qualitatively similar results to the one for the von Neumann entanglement entropy. Figure~\ref{fig6}(a) shows that, for the Dirac SYK2 model~(\ref{dsyk2}), the differences decrease with increasing system size. The scaling observed suggests that the first subleading correction to the leading volume-law term is a negative constant. The results in Fig.~\ref{fig6}(b), for the Majorana SYK2 model~(\ref{msyk2}) in the main panel and for the general quadratic model~(\ref{eqGQ}) in the inset, are qualitatively similar. The main difference is that the subleading constant term is positive.

\begin{figure}[!t]
\includegraphics[width=0.95\columnwidth]{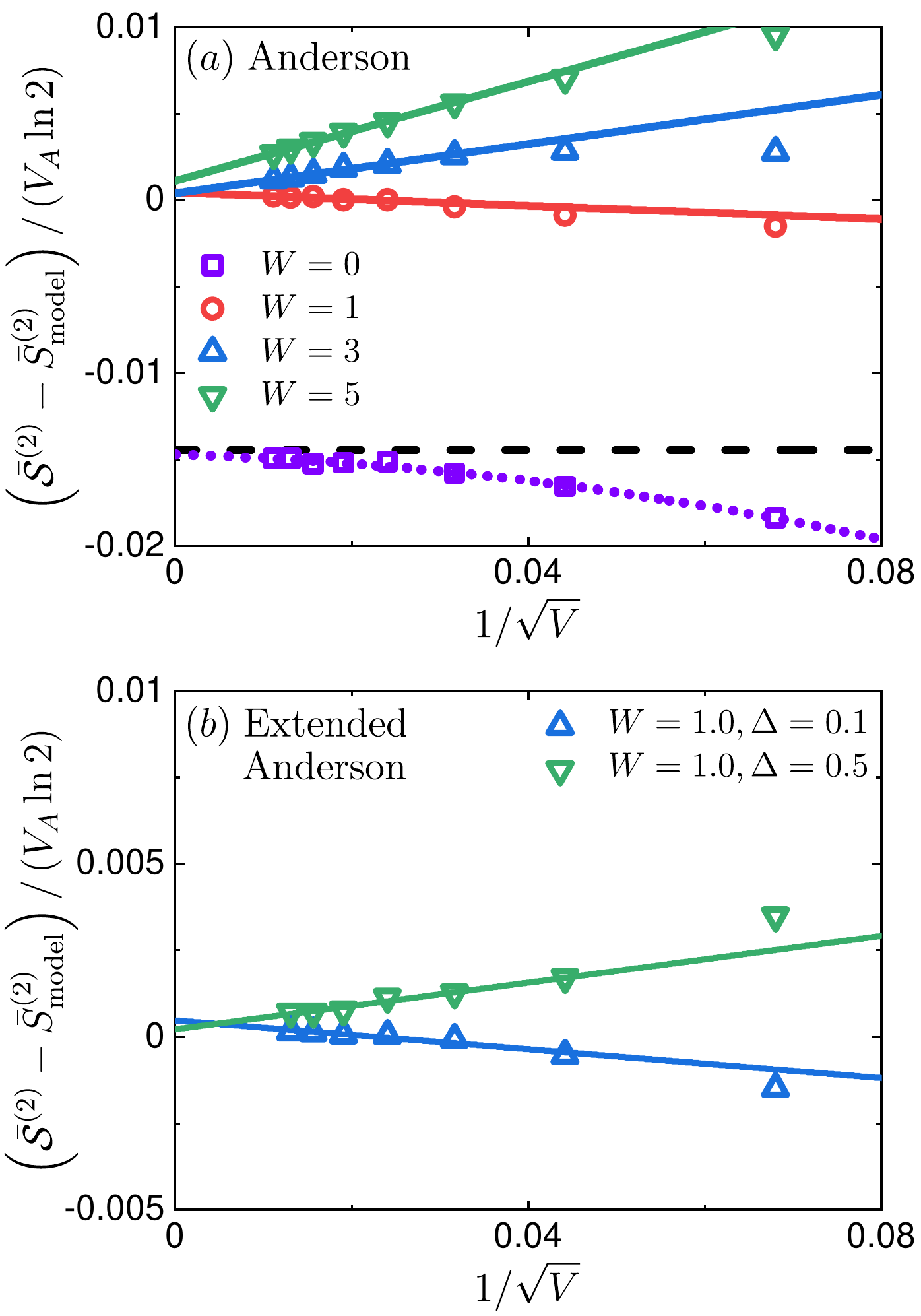}
\vspace{-0.1cm}
\caption{Average second R{\'e}nyi entanglement entropies $\bar S_{\rm model}^{(2)}$ of eigenstates of: (a) the 3D Anderson model~(\ref{eqAn}) and (b) the extended 3D Anderson model~(\ref{eqH}) at subsystem fraction $f=1/2$. The averages were computed as in Fig.~\ref{fig5}. The solid lines show the results of two-parameter fits to $a_0+a_1/\sqrt{V}$ using the data for $V \geq 512$. We get $a_0\leq 5\cdot 10^{-4}$ for $W = 1$ and 3, and $a_0 = 10^{-3}$ for $W = 5$. In (a), the dotted line corresponds to a three-parameter fit to $a_0+a_1/\sqrt{V}+a_2/V$ using the data for $V \geq 512$ at $W=0$.
The horizontal dashed line marks the thermodynamic limit result obtained numerically for translationally invariant noninteracting fermions in one dimension, $\bar{S}_{W=0}^{(2)}/(V_A\ln 2)\approx 0.4713$, see Appendix~\ref{sec:app2}.}
\label{fig7}
\end{figure}

Next, we compare $\bar{\mathcal{S}}^{(2)}$ from Eq.~$\left(\ref{eqS}\right)$ with numerical results for the average second R{\'e}nyi entanglement entropy $\bar S^{(2)}$ of eigenstates of the local quantum-chaotic quadratic Hamiltonians (3D Anderson models) introduced in Sec.~\ref{subsec:local} at $f=1/2$. [In Fig.~\ref{fig1} we already showed that, for the particle-number conserving 3D Anderson model~(\ref{eqAn}) in a lattice with $V = 8000$ sites and $W=1$, the numerical results for $\bar S^{(2)}$ as a function of the subsystem fraction $f$ are accurately described by the analytical prediction $\bar{\mathcal{S}}^{(2)}$ from Eq.~\eqref{eqS}.] The results in Fig.~\ref{fig7}(a) for the particle-number conserving 3D Anderson model~(\ref{eqAn}) in the delocalized regime, and in Fig.~\ref{fig7}(b) for the extended (particle-number nonconserving) Anderson model~(\ref{eqH}), are qualitatively similar to the ones for the von Neumann entanglement entropy shown in Figs.~\ref{fig5}(a) and~\ref{fig5}(b), respectively.

For the particle-number conserving model in Fig.~\ref{fig7}(a), as mentioned before, $W=0$ corresponds to a translationally invariant model, which does not exhibit single-particle quantum chaos. In that case the results appear to converge, with increasing system size, to the thermodynamic limit result obtained numerically for noninteracting fermions in one dimension, $\lim_{V,V_A \to\infty}\bar S^{(2)}/(V_A \ln 2) \approx 0.4713$ at $f=1/2$ [see the horizontal dashed line in Fig.~\ref{fig7}(a), and Appendix~\ref{sec:app2}]. We note that it is larger than the value predicted by Eq.~\eqref{eq:s2f2} for quantum-chaotic quadratic Hamiltonians. It is interesting that, in contrast, the volume-law coefficient of the average von Neumann entanglement entropy of translationally-invariant noninteracting fermions is smaller than the one for quantum-chaotic quadratic Hamiltonians, see Fig.~\ref{fig5}(a).

\section{Summary and discussion} \label{sec:conclusions}

We studied the entanglement properties of many-body eigenstates of various quantum-chaotic quadratic Hamiltonians, namely, of quadratic Hamiltonians that exhibit single-particle quantum chaos. By the latter we mean that the statistics of the single-particle energy levels is described by the random matrix theory. We argued that the entanglement entropies of typical many-body eigenstates of those Hamiltonians exhibit universal features. Specifically, we reported numerical evidence that the analytical expression for the average von Neumann entanglement entropy introduced in Ref.~\cite{lydzba_rigol_20}, and for the second R{\'e}nyi entanglement entropy introduced here, describe the results for exemplary quantum-chaotic quadratic Hamiltonians. We considered both local and nonlocal Hamiltonians, as well as particle-number conserving and nonconserving versions of each of them. We also showed that the average entanglement entropies of many-body eigenstates of local translationally-invariant quadratic Hamiltonians, which do not exhibit quantum chaos but have single-particle eigenstates that are delocalized in real space, exhibit small but robust deviations from the analytical predictions for quantum-chaotic quadratic Hamiltonians.

Several questions are left open for future work. Among those is whether there is a close-form expression for the average second R{\'e}nyi entanglement entropy in Eq.~(\ref{eqS}), for which we provided closed-form expressions in the limits of subsystem fractions $f\rightarrow 0$ and $f=1/2$. Also, our analytical treatment focused on the leading (volume law) term in the entanglement entropies. As discussed in Secs.~\ref{sec:vN} and~\ref{sec:Renyi}, the subleading terms exhibit interesting behaviors that still need to be understood analytically. Finally, as shown in Fig.~\ref{fig1}, we note that the averages (over all eigenstates) of both the von Neumann and second R{\'e}nyi entanglement entropies are concave functions of $f$. What happens in ``finite-temperature'' averages is an open question. For eigenstates of generic Hamiltonians, a qualitative difference between the von Neumann and second R{\'e}nyi entanglement entropies was conjectured to emerge for averages in microcanonical windows away from infinite temperature~\cite{lu_grover_19}: while the von Neumann entropy is expected to increase linearly with $f$, the second (and higher) R{\'e}nyi entropies were conjectured to be convex functions. 

\acknowledgements
This work was supported by the the Slovenian Research Agency (ARRS), Research core fundings Grants No.~P1-0044 and No.~J1-1696 (P.\L.~and L.V.) and by the National Science Foundation, Grant No.~2012145 (M.R.).

\appendix

\section{Fluctuations over Hamiltonian realizations}
\label{sec:app1}

\begin{figure}[!t]
\includegraphics[width=0.95\columnwidth]{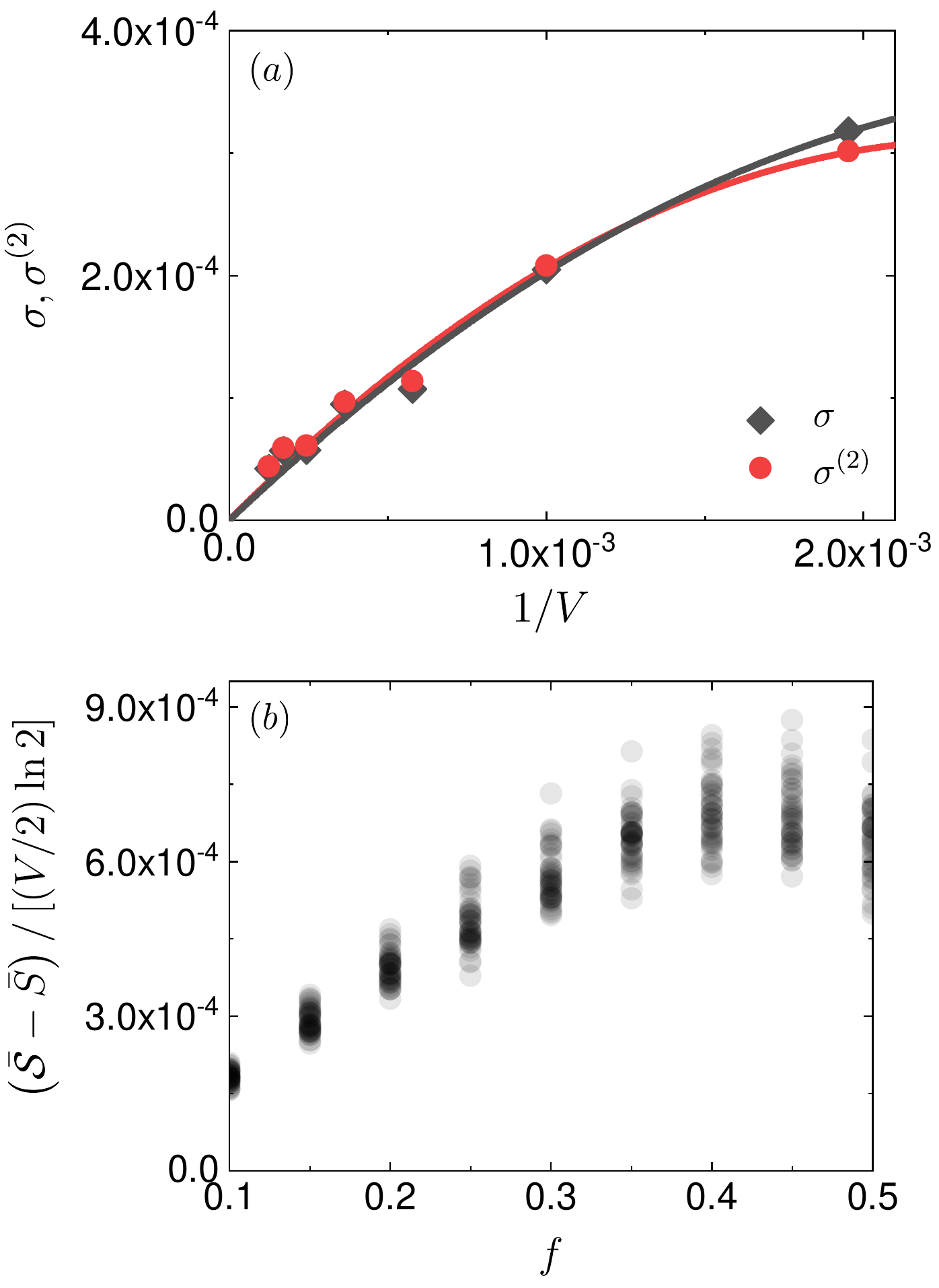}
\vspace{-0.1cm}
\caption{(a) Realization-to-realization standard deviations $\sigma$ and $\sigma^{(2)}$, see Eq.~(\ref{def_sigma}), versus $1/V$ for the 3D Anderson model with $W=1$. We consider systems with $V\in\left\{8^3,10^3,...,20^3\right\}$, at subsystem fraction $f=1/2$. The lines show results of second-order polynomial fits to $a_1/V-a_2/V^2$, with fitting parameters $a_1$ and $a_2$. (b) Relative differences between the analytical prediction $\bar{\mathcal{S}}$ and numerical results $\bar{S}$ for the von Neumann entanglement entropy, versus $f$. Each point depicts the result obtained for a single disorder realization in a system with $V=20^3$. For each $f$, we report results for $50$ different disorder realizations. Note that all points are light gray so darker means a higher number of overlapping points.}\label{figA1}
\end{figure}

In the main text, we compared analytical results for $\bar{\mathcal{S}}$ and $\bar{\mathcal{S}}^{(2)}$ [cf.~Eqs.~(\ref{eqAS}) and~(\ref{eqS})] with numerical results $\bar{{S}}$ and $\bar{{S}}^{(2)}$ for the von Neumann and second R{\'e}nyi entanglement entropies, respectively. The numerical results were obtained by first averaging over Hamiltonian eigenstates for a single Hamiltonian realization, and then averaging over different Hamiltonian realizations.

Here we explore the realization-to-realization fluctuations. To this end, we compute $\bar{S}$ as an eigenstate average for a single Hamiltonian realization, and define $\left<...\right>$ as the average over different Hamiltonian realizations. We define the corresponding realization-to-realization standard deviation for the von Neumann entanglement entropy as
\begin{equation} \label{def_sigma}
\sigma=\frac{\sqrt{\left<\bar{S}^2\right>-\left<\bar{S}\right>^2}}{V_A\ln 2}\,.
\end{equation}
In what follows $\bar{S}$ corresponds to the average over $500$ random many-body eigenstates in a single Hamiltonian, while the average $\left<...\right>$ is computed over $50$ Hamiltonian realizations. We also compute the realization-to-realization standard deviation for the second R{\'e}nyi entanglement entropy $\sigma^{(2)}$, for which $\bar{S}$ in Eq.~(\ref{def_sigma}) is replaced by $\bar{S}^{(2)}$.

Figure~\ref{figA1}(a) shows the standard deviations $\sigma$ and $\sigma^{(2)}$ for the 3D Anderson model at $W=1$ for different system sizes. $\sigma$ and $\sigma^{(2)}$ are very small (of the order $10^{-4}$) for the system sizes considered in the main text, and appear to vanish with increasing system size. In Fig.~\ref{figA1}(b), we plot the relative differences $(\bar{\mathcal{S}} - \bar S)/[(V/2) \ln2]$ for $50$ different disorder realizations, as a function of the subsystem fraction $f$. These results show that the realization-to-realization fluctuations are smaller than the finite-size corrections to the analytic result $\bar{\mathcal{S}}$ in the thermodynamic limit. Similar results (not shown) were obtained for the second R{\'e}nyi entanglement entropy.

\section{Second R{\'e}nyi entropy for translationally invariant free fermions}
\label{sec:app2}

\begin{figure}[!t]
\includegraphics[width=0.95\columnwidth]{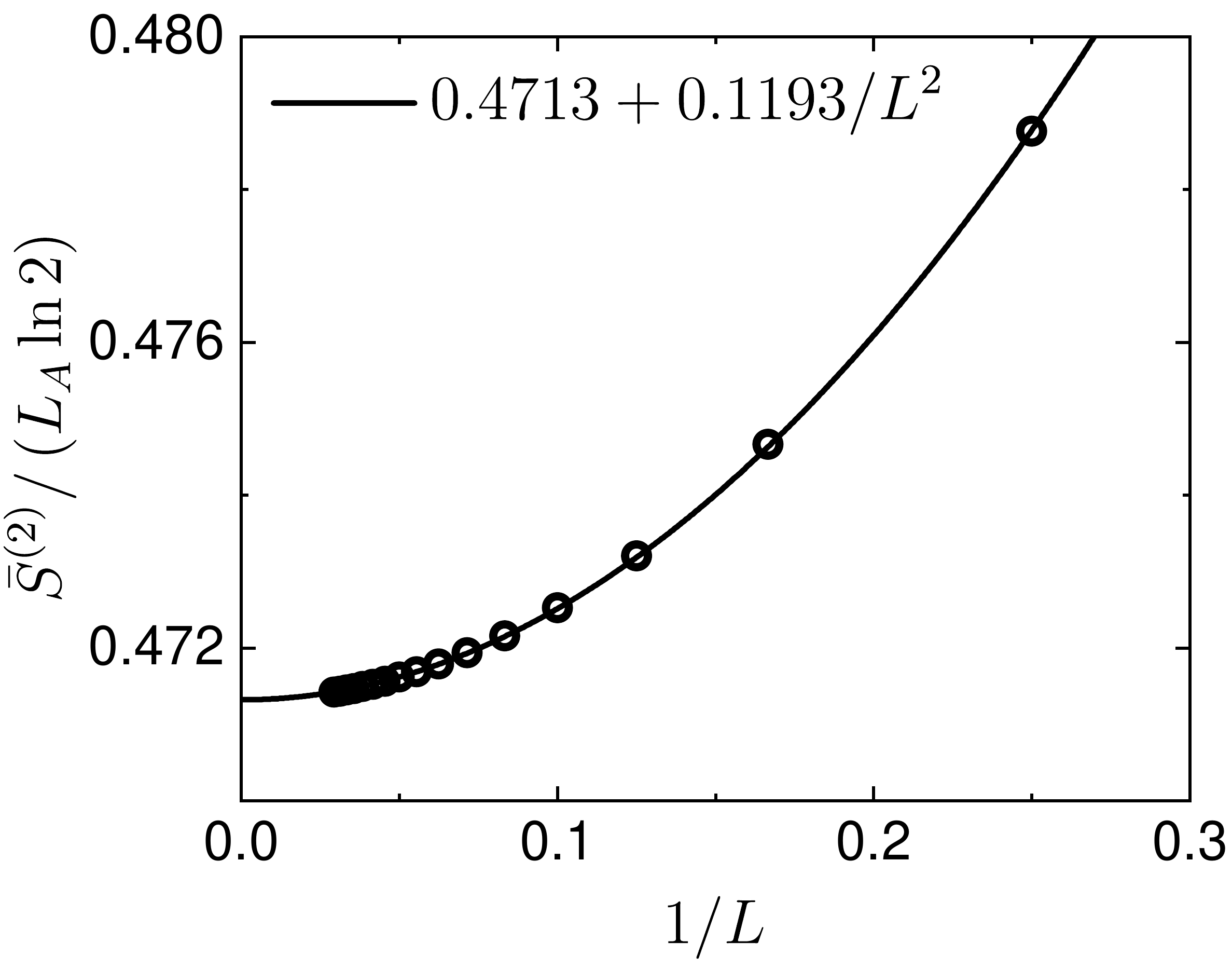}
\vspace{-0.1cm}
\caption{The average second R{\'e}nyi entanglement entropy $\bar{S}^{(2)}/(L_A\ln 2)$ versus $1/L$, at $f=1/2$, for translationally invariant free fermions in 1D.
The solid line shows the result of a quadratic fit $a_0+a_2/L^2$ to the data, yielding $a_0\approx 0.4713$.}
\label{figA2}
\end{figure}

In Fig.~\ref{fig7}(a), we show results for the average second R{\'e}nyi entanglement entropy for $W=0$, i.e., for translationally invariant noninteracting fermions in three dimensions. Those results are compared to the result obtained for translationally invariant noninteracting fermions in one dimension (1D) [horizontal dashed line in Fig.~\ref{fig7}(a)]. Here we show how the latter result was obtained.

We calculated the average second R{\'e}nyi entanglement entropy $\bar S^{(2)}$ for translationally invariant free fermions in 1D using Eq.~(\ref{eq:aree}) for systems with linear sizes up to $L = 34$ (we set $V \to L$ and $V_A \to L_A$), so that the average can be computed over all $2^L$ eigenstates. Results for subsystem fraction $f=1/2$ are shown in Fig.~\ref{figA2} as a function of $1/L$. The dominant subleading contribution to $\bar{S}^{(2)}/(L_A\ln 2)$ scales as $\propto 1/L^2$. Using a fit to a quadratic function, we obtain an accurate estimate for the volume-law coefficient in the thermodynamic limit, $\lim_{L,L_A \to \infty}\bar{S}^{(2)}/(L_A\ln 2) \approx 0.4713$. It is interesting to note the different scalings of the dominant subleading terms in the translationally invariant free fermion Hamiltonian in 1D. While for the average second R{\'e}nyi entanglement shown here it vanishes as a power law, it vanishes exponentially fast with $L$ for the average von Neumann entanglement entropy~\cite{vidmar_hackl_17}. 

\bibliographystyle{biblev1}
\bibliography{references}

\end{document}